\documentclass[a4paper,12pt]{article}%
\usepackage{amsmath}%
\usepackage{amsfonts}%
\usepackage{amssymb}%
\usepackage{graphicx}

\begin{document}

\title{SWAP OPERATORS AND ELECTRIC CHARGES OF FERMIONS}
\maketitle
\begin{center}
\author{\footnote{D\'{e}partement d'Ing\'{e}nierie Civil,
Institut Sup\'{e}rieur de Technologie
d'Antananarivo, 
IST-T, BP 8122, Madagascar.\\
D\'{e}partement de Physique, Laboratoire de Rh\'{e}ologie des Suspensions, LRS, Universit\'{e} d'Antananarivo, Madagascar.\\ E-mail: rakotopierre@refer.mg\\}Christian Rakotonirina and \footnote{D\'{e}partement de Physique,
Laboratoire de Rh\'{e}ologie des Suspensions, LRS, Universit\'{e} d'Antananarivo,
Madagascar.\\E-mail: aaratiarison@yahoo.fr}Adolphe A. Ratiarison}
\end{center}

\begin{abstract}
 \noindent  Electric Charges operator (ECO) in phase space formulation, proposed by Zenczykowski, is expressed
in terms of a swap operator, in some expressions for possible physical interpretations. An expression
of an ECO in terms of a swap operator makes sense to the eigenvalues of the swap operator. An ECO
including all the fermions of the standard model (SM) is constructed. 
\end{abstract}

\section{Introduction}

Swap operator or tensor commutation matrix $n\otimes n$ is the matrix $U_{n\otimes n}$ which has the following properties: for any unicolumn and n rows matrices $a \in\mathbb{C}^{n\times1}$, $b\in\mathbb{C}^{n\times1}$ 
\begin{equation*}
U_{n\otimes n}\cdot\left(a\otimes b\right)=b\otimes a
\end{equation*}
and for any two $n\times n$-matrices, $A$,
$B$ $\in\mathbb{C}^{n\times n}$
\begin{equation}\label{eq3}
U_{n\otimes n}\cdot\left(A\otimes B\right) = \left(B\otimes A\right)\cdot U_{n\otimes n}
\end{equation}
In quantum information theory (Cf. for example \cite{ref1}) the swap operator $2\otimes 2$
\begin{equation*}
U_{2\otimes2}=
\begin{pmatrix}
  1 & 0 & 0 & 0 \\
  0 & 0 & 1 & 0 \\
  0 & 1 & 0 & 0 \\
  0 & 0 & 0 & 1 \\
\end{pmatrix}
\end{equation*}
 is written as linear combination of the tensor products of the Pauli matrices $\sigma_1=\begin{pmatrix}
  0 & 1 \\
  1 & 0 \\
\end{pmatrix}$, $\sigma_2=\begin{pmatrix}
  0 & -i \\
  i & 0 \\
\end{pmatrix}$, $\sigma_3=\begin{pmatrix}
  1 & 0 \\
  0 & -1 \\
\end{pmatrix}$ and the unit $2\times2$-matrix $\sigma_0=\begin{pmatrix}
  1 & 0 \\
  0 & 1 \\
\end{pmatrix}$.
\begin{equation}\label{eq4}
U_{2\otimes2}=\frac{1}{2}\sigma_0\otimes\sigma_0+\frac{1}{2}\sum_{i=1}^3\sigma_i\otimes \sigma_i
\end{equation}

Since $U_{2\otimes2}$ is an unitary matrix, it gives a representation of the Dirac equation. \cite{ref2} thinks to have given physical interpretation of the transformed 

\begin{equation*}
U_{2\otimes2}\cdot\left(\sqrt{\frac{E+mc^2}{2E}}\frac{1}{\sqrt{2\left(1+n_3\right)}}e^{\frac{i}{\hbar}\left(\vec{p}\cdot\vec{x}-Et\right)}\begin{pmatrix}
  1 \\
  \frac{cp}{E+mc^2} \\
\end{pmatrix}\otimes \begin{pmatrix}
  1+n_3 \\
  n_1+in_2 \\
\end{pmatrix}\right)
\end{equation*}
\begin{equation*}
=\sqrt{\frac{E+mc^2}{2E}}\frac{1}{\sqrt{2\left(1+n_3\right)}}e^{\frac{i}{\hbar}\left(\vec{p}\cdot\vec{x}-Et\right)}\begin{pmatrix}
  1+n_3 \\
  n_1+in_2 \\
\end{pmatrix}\otimes \begin{pmatrix}
  1 \\
  \frac{cp}{E+mc^2} \\
\end{pmatrix}
\end{equation*}
by $U_{2\otimes2}$ of a solution of the Dirac equation.

Fermions have quantum numbers $I_3$, the isospin and $Y$, the hypercharge. The electric charge $Q$ of a fermion is given by the Gell-Mann-Nishijima relation

\begin{equation*}
Q=I_3+\frac{Y}{2}
\end{equation*}

\noindent For the fermions of the SM these quantum numbers are given by the following table.

\begin{center}
	
	\begin{tabular}{l|r|r|rrr}
		     &   & $Q$ & $I_3$ & $Y$\\
		\hline
		  Neutral leptons & $\nu_{eL}$, $\nu_{\mu L}$, $\nu_{\tau L}$ & $0$ & $1/2$ & $-1$\\
		 
		\hline
	  Charged leptons & $e_L$, $\mu_L$, $\tau_L$ & $-1$ & $-1/2$ & $-1$\\
		   & $e_R$, $\mu_R$, $\tau_R$ & $-1$ & $0$ & $-2$ \\
		\hline
	  Quarks $u$, $c$, $t$ & $u^{r}_L$, $u^{b}_L$, $u^{g}_L$, $c^{r}_L$, $c^{b}_L$, $c^{g}_L$, $t^{r}_L$, $t^{b}_L$, $t^{g}_L$ & $2/3$ & $1/2$ & $1/3$\\
		   & $u^{r}_R$, $u^{b}_R$, $u^{g}_R$, $c^{r}_R$, $c^{b}_R$, $c^{g}_R$, $t^{r}_R$, $t^{b}_R$, $t^{g}_R$ & $2/3$ & $0$ & $4/3$\\
		
	 \hline
	  Quarks $d$, $s$, $b$ & $d^{r}_L$, $d^{b}_L$, $d^{g}_L$, $s^{r}_L$, $s^{b}_L$, $s^{g}_L$, $b^{r}_L$, $b^{b}_L$, $b^{g}_L$ & $-1/3$ & $-1/2$ & $1/3$\\
		   & $d^{r}_R$, $d^{b}_R$, $d^{g}_R$, $s^{r}_R$, $s^{b}_R$, $s^{g}_R$, $b^{r}_R$, $b^{b}_R$, $b^{g}_R$ & $-1/3$ & $0$ & $-2/3$\\
	\hline
						
		\end{tabular}						
\end{center}
 
A matrix Gell-Mann-Nishijima relation for eight leptons and quarks of the SM of the same generation is proposed by \cite{ref3}, in phase space formulation. According to the formula (\ref{eq4}) it is easy to notice that this matrix Gell-Mann-Nishijima relation can be expressed in terms of $U_{2\otimes2}$. In this paper, we will write this relation in some forms where physical interpretation of the action of the swap operator $U_{2\otimes2}$ is possible. Then the eigenvalues and eigenvectors of $U_{2\otimes2}$ will take physical sense. In the section $4$, for including more fermions of the SM we will write a matrix formula giving the electric charges in terms of the swap operator $3\otimes3$,

\begin{equation*}
U_{3\otimes3}=
\begin{pmatrix}
  1 & 0 & 0 & 0 & 0 & 0 & 0 & 0 & 0 \\
  0 & 0 & 0 & 1 & 0 & 0 & 0 & 0 & 0 \\
  0 & 0 & 0 & 0 & 0 & 0 & 1 & 0 & 0 \\
  0 & 1 & 0 & 0 & 0 & 0 & 0 & 0 & 0 \\
  0 & 0 & 0 & 0 & 1 & 0 & 0 & 0 & 0 \\
  0 & 0 & 0 & 0 & 0 & 0 & 0 & 1 & 0 \\
  0 & 0 & 1 & 0 & 0 & 0 & 0 & 0 & 0 \\
  0 & 0 & 0 & 0 & 0 & 1 & 0 & 0 & 0 \\
  0 & 0 & 0 & 0 & 0 & 0 & 0 & 0 & 1 \\
\end{pmatrix}
\end{equation*}
Then the physical sense given to the eigenvalues of the swap operators are maintained. In section $5$, for including all the fermions of the SM we will write the matrix of the electric charges in terms of the swap operator $U_{4\otimes4}$. For the calculus we have used SCILAB, a free mathematical software for numerical analysis.

\section{Gell-Mann-Nishijima relation in phase space formulation}
We re-write here the Gell-Mann-Nishijima relation in phase space approch \cite{ref3}, which yields the ECO of eight fermions, two leptons and two three-colored quarks for a single SM generation, for example $e_L$, $\nu_{eL}$, $u^{r}_L$, $u^{b}_L$, $u^{g}_L$, $d^{r}_L$, $d^{b}_L$, and $d^{g}_L$. 

\begin{equation}\label{eq6}
\textbf{Q}=\textbf{I}_3+\frac{\textbf{Y}}{2}
\end{equation} 
where 
\begin{equation*}
\textbf{I}_3=\frac{1}{2}\sigma_0\otimes\sigma_0\otimes\sigma_3 
\end{equation*} 
the weak isospin, 
\begin{equation*}
\textbf{Y}=\left(\frac{1}{3}\sum_{i=1}^3\sigma_i\otimes \sigma_i\right)\otimes \sigma_0
\end{equation*} 
the weak hypercharge.\\

\noindent We remark that the operators $\textbf{I}_3$ and $\textbf{Y}$ act independantly on the vector field because in the expression of $\textbf{I}_3$, $\sigma_3$ is at the right side of the tensor product and in the expression of $\textbf{Y}$,  $\left(\frac{1}{3}\displaystyle\sum_{i=1}^3\sigma_i\otimes \sigma_i\right)$ is at the left side.

\section{Electric charges operator in terms of the swap Operator}

According to the formula (\ref{eq4}) we can introduce the operator $U_{2\otimes2}$ into the expression of the operator weak hypercharge $\textbf{Y}$, and then into the expression of the ECO $\textbf{Q}$.

\begin{equation*}
\textbf{Y}=\frac{2}{3}U_{2\otimes2}\otimes \sigma_0-\frac{1}{3}\sigma_0\otimes\sigma_0\otimes\sigma_0
\end{equation*}
Hence
\begin{equation}\label{eq10}
\textbf{Q}=\frac{1}{2}\sigma_0\otimes\sigma_0\otimes\sigma_3+\frac{1}{3}\left(U_{2\otimes2}\otimes \sigma_0-\frac{1}{2}\sigma_0\otimes\sigma_0\otimes\sigma_0\right)
\end{equation}
Or
\begin{equation*}
\textbf{Q}=\sigma_0\otimes\sigma_0\otimes\left(\frac{\sigma_3}{2}+\alpha\frac{\sigma_0}{6}\right)+\frac{1}{3}\left(U_{2\otimes2} -\frac{1+\alpha}{2}\sigma_0\otimes\sigma_0\right)\otimes\sigma_0
\end{equation*}
with $\alpha$ a real parameter.

\noindent If $\alpha=0$, we have the relation (\ref{eq10}), that is (\ref{eq6}).\\

\noindent If $\alpha=-3$,
\begin{equation}\label{eq12}
\textbf{Q}=\sigma_0\otimes\sigma_0\otimes Q_L+\frac{1}{3}\left(U_{2\otimes2}+\sigma_0\otimes\sigma_0\right)\otimes\sigma_0
\end{equation}
 
\noindent with $Q_L=\begin{pmatrix}
  0 & 0 \\
  0 & -1 \\
\end{pmatrix}$ whose diagonal is formed by the electric charges of $\nu_{eL}$ and $e_L$.\\
 
\noindent If $\alpha=1$, 
\begin{equation}\label{eq13}
\textbf{Q}=\sigma_0\otimes\sigma_0\otimes Q_Q+\frac{1}{3}\left(U_{2\otimes2}-\sigma_0\otimes\sigma_0\right)\otimes\sigma_0
\end{equation} 
with $Q_Q=\begin{pmatrix}
  2/3 & 0 \\
  0 & -1/3 \\
\end{pmatrix}$ whose diagonal is formed by the charges of a up quark $u$ and a down quark $d$.\\

\noindent The eigenvalues of $U_{2\otimes2}$ are once $-1$ and three times $+1$.\\

\noindent $\frac{1}{\sqrt{2}}\begin{pmatrix}
  0 \\
  -1 \\
  1\\
  0
\end{pmatrix}$ the eigenvector of $U_{2\otimes2}$ associated to the eigeinvalue $-1$, which is, according to (\ref{eq12}), associated to leptons. \\
$\begin{pmatrix}
  1 \\
  0 \\
  0\\
  0
\end{pmatrix}$, $\frac{1}{\sqrt{2}}\begin{pmatrix}
  0 \\
  1 \\
  1\\
  0
\end{pmatrix}$, $\begin{pmatrix}
  0 \\
  0 \\
  0\\
  1
\end{pmatrix}$ are the eigeinvectors associated to $+1$, which are, according to (\ref{eq13}), associated to three-colored quarks.\\

The number of the eigenvalue $-1$ of $U_{2\otimes2}$ are the number of the generation of leptons. $+1$ three times the eigenvalue of $U_{2\otimes2}$, that is the number of the colors. Hence the eigenvalues of $\textbf{Q}$ are the charges of the eight fermions mentioned above.  

\section{Electric Charges Operator in Terms of Swap Operator $3\otimes3$}
Let, for example, $Q_L=\begin{pmatrix}
  0 & 0 & 0  \\
  0 & -1 & 0  \\
  0 & 0 & -1  \\
\end{pmatrix}$ whose diagonal is formed by the electric charge of a neutrinos and electric charges of two charged leptons and $Q_Q=\begin{pmatrix}
  2/3 & 0 & 0  \\
  0 & -1/3 & 0  \\
  0 & 0 & -1/3  \\
\end{pmatrix}$ 
whose diagonal is formed by the electric charge of a up quark u (a charm quark c or a top quark t) and electric charges of down quarks, strange quarks or bottom quarks.
\begin{equation}\label{eq13prime}
Q_Q-Q_L=\frac{2}{3}\lambda_0
\end{equation}
where $\lambda_0$ is the $3\times 3$-unit matrix.\\
Hence, 
\begin{equation*}
\lambda_0\otimes \lambda_0\otimes Q_Q + \frac{1}{3}(U_{3\otimes 3}-\lambda_0\otimes \lambda_0)\otimes \lambda_0=\lambda_0\otimes \lambda_0\otimes Q_L + \frac{1}{3}(U_{3\otimes 3}+\lambda_0\otimes \lambda_0)\otimes \lambda_0 
\end{equation*}

\noindent We denote it $\textbf{Q}$ like electric charge operator. \\
The eigenvalues of the swap operator $U_{3\otimes 3}$ are $-1$ three times and $+1$ six times. From the above equation the eigenvalues $-1$ are associated to leptons wheares the eigenvalues $+1$ are associated to quarks. Following the result of \cite{ref3}, the three eigenvalues $-1$ are associated to the three generations of leptons, whereas the three eigenvalues $+1$ are associated to three colors of left handed quarks and the three ones are associated to three colors of right handed quarks. The diagonal of $Q_L$ are formed by the charges of the leptons of the SM in a same generation, for example $\nu_{eL}$, $e_L$ and $e_R$. The diagonal of $Q_Q$ are formed by the charges of a up quark $u$, a down quark $d$ and a strange quark $s$. Hence the twenty seven eigenvalues of $\textbf{Q}$, $-1$ six times, $0$ three times, $-1/3$ twelve times and $+2/3$ six times can be the charges of following SM fermions $\nu_{eL}$, $\nu_{\mu L}$, $\nu_{\tau L}$,  $e_L$, $\mu_L$, $\tau_L$, $e_R$, $\mu_R$, $\tau_R$,  $u^{r}_L$, $u^{b}_L$, $u^{g}_L$, $u^{r}_R$, $u^{b}_R$, $u^{g}_R$,  $d^{r}_L$, $d^{b}_L$, $d^{g}_L$, $d^{r}_R$, $d^{b}_R$, $d^{g}_R$, $s^{r}_L$, $s^{b}_L$, $s^{g}_L$, $s^{r}_R$, $s^{b}_R$, $s^{g}_R$. 
\begin{equation*}
Q_Q=\frac{1}{2}\lambda_3+\frac{1}{2\sqrt{3}}\lambda_8 
\end{equation*}
with\\
 $\lambda_1=\begin{pmatrix}
  0 & 1 & 0  \\
  1 & 0 & 0  \\
  0 & 0 & 0  \\
\end{pmatrix}$, $\lambda_2=\begin{pmatrix}
  0 & -i & 0  \\
  i & 0 & 0  \\
  0 & 0 & 0  \\
\end{pmatrix}$, $\lambda_3=\begin{pmatrix}
  1 & 0 & 0  \\
  0 & -1 & 0  \\
  0 & 0 & 0  \\
\end{pmatrix}$,
 $\lambda_4=\begin{pmatrix}
  0 & 0 & 1  \\
  0 & 0 & 0  \\
  1 & 0 & 0  \\
\end{pmatrix}$, 
$\lambda_5=\begin{pmatrix}
  0 & 0 & -i  \\
  0 & 0 & 0  \\
  i & 0 & 0  \\
\end{pmatrix}$, 
$\lambda_6=\begin{pmatrix}
  0 & 0 & 0  \\
  0 & 0 & 1  \\
  0 & 1 & 0  \\
\end{pmatrix}$, $\lambda_7=\begin{pmatrix}
  0 & 0 & 0  \\
  0 & 0 & -i  \\
  0 & i & 0  \\
\end{pmatrix}$, 
$\lambda_8=\frac{1}{\sqrt{3}}\begin{pmatrix}
  1 & 0 & 0  \\
  0 & 1 & 0  \\
  0 & 0 & -2  \\
\end{pmatrix}$ \\
are the Gell-Mann matrices.\\

The swap operator $U_{3\otimes 3}$ can be written in terms of the Gell-Mann matrices under the following way \cite{ref4}
\begin{equation*}
U_{3\otimes 3}=\frac{1}{3}\lambda_0\otimes \lambda_0+\frac{1}{2}\sum_{i=1}^8\lambda_i\otimes\lambda_i 
\end{equation*}

\noindent So,
\begin{equation*}
\textbf{Q}=\lambda_0\otimes \lambda_0\otimes \left(-\frac{2}{9}\lambda_0+\frac{1}{2}\lambda_3+\frac{1}{2\sqrt{3}}\lambda_8\right)+\frac{1}{6}\left(\sum_{i=1}^8\lambda_i\otimes\lambda_i\right)\otimes \lambda_0
\end{equation*}

\noindent \textbf{Q} can be written under the form of the relation (\ref{eq6}), where
\begin{equation*}
 \textbf{I}_3=\frac{1}{2}\left(\lambda_0\otimes\lambda_0\otimes\lambda_3-\tau_1\otimes\tau_1\otimes\tau_1-\tau_2\otimes\tau_2\otimes\tau_1-\tau_3\otimes\tau_3\otimes\tau_1\right)
 \end{equation*}

\begin{equation*}
 \textbf{Y}=\tau_1\otimes\tau_1\otimes\tau_1+\tau_2\otimes\tau_2\otimes\tau_1+\tau_3\otimes\tau_3\otimes\tau_1+\frac{1}{\sqrt{3}}\lambda_0\otimes\lambda_0\otimes\lambda_8+\frac{2}{3}\left(U_{3\otimes 3}-\lambda_0\otimes\lambda_0\right)\otimes\lambda_0
 \end{equation*}
 \noindent or
 
 \begin{equation*}
 \textbf{Y}=\tau_1\otimes\tau_1\otimes\tau_1+\tau_2\otimes\tau_2\otimes\tau_1+\tau_3\otimes\tau_3\otimes\tau_1+\lambda_0\otimes\lambda_0\otimes\left(-\frac{4}{9}\lambda_0+\frac{1}{\sqrt{3}}\lambda_8\right)+\frac{1}{3}\left(\sum_{i=1}^8\lambda_i\otimes\lambda_i\right)\otimes \lambda_0
 \end{equation*}
 
 \noindent with $\tau_1=\begin{pmatrix}
  1 & 0 & 0  \\
  0 & 0 & 0  \\
  0 & 0 & 0  \\
\end{pmatrix}$, $\tau_2=\begin{pmatrix}
  0 & 0 & 0  \\
  0 & 1 & 0  \\
  0 & 0 & 0  \\
\end{pmatrix}$ and $\tau_3=\begin{pmatrix}
  0 & 0 & 0  \\
  0 & 0 & 0  \\
  0 & 0 & 1  \\
\end{pmatrix}$.\\

\noindent From (\ref{eq3}), $\textbf{I}_3$ and $\textbf{Y}$ commute, so they are simultaneously diagonalizables.\\
 
 If we take $Q_L=\begin{pmatrix}
  0 & 0 & 0  \\
  0 & 0 & 0  \\
  0 & 0 & -1  \\
\end{pmatrix}$ and $Q_Q=\begin{pmatrix}
  2/3 & 0 & 0  \\
  0 & 2/3 & 0  \\
  0 & 0 & -1/3  \\
\end{pmatrix}$ 
the above relations between $Q_L$ and $Q_Q$ will hold. The twenty seven eigenvalues of the ECO $\textbf{Q}$ are $-1$ three times, $0$ six times, $-1/3$ six times and $+2/3$ twelve times. These eigenvalues can be the charges of the following  fermions $\nu_{eL}$, $\nu_{\mu L}$, $\nu_{\tau L}$,  $\nu_{eR}$, $\nu_{\mu R}$, $\nu_{\tau R}$, $e_L$, $\mu_L$, $\tau_L$, $c^{r}_L$, $c^{b}_L$, $c^{g}_L$, $c^{r}_R$, $c^{b}_R$, $c^{g}_R$,  $t^{r}_L$, $t^{b}_L$, $t^{g}_L$, $t^{r}_R$, $t^{b}_R$, $t^{g}_R$, $b^{r}_L$, $b^{b}_L$, $b^{g}_L$, $b^{r}_R$, $b^{b}_R$, $b^{g}_R$. The right handed neutrinos $\nu_{eR}$, $\nu_{\mu R}$, $\nu_{\tau R}$ whose charge is $0$ are not SM fermions.

\section{Including all the fermions of the standard Model}
For including all the fermions of the SM we are going to build an ECO in terms of the swap operator $U_{4\otimes 4}$.\\
The eigenvalues of the swap operator $U_{4\otimes 4}$ are $-1$ six times and $+1$ ten times. 
So, the ECO   
\begin{equation*}
\textbf{Q}=\Lambda_0\otimes \Lambda_0\otimes Q_Q + \frac{1}{3}(U_{4\otimes 4}-\Lambda_0\otimes \Lambda_0)\otimes \Lambda_0=\Lambda_0\otimes \Lambda_0\otimes Q_L + \frac{1}{3}(U_{4\otimes 4}+\Lambda_0\otimes \Lambda_0)\otimes \Lambda_0 
\end{equation*}
where $\Lambda_0$ is the $4\times 4$-unit matrix and $Q_Q=\begin{pmatrix} 
  -1/3 & 0 & 0 & 0 \\
  0 & 2/3 & 0 & 0 \\
  0 & 0 & -1/3 & 0 \\
  0 & 0 & 0 & 2/3 \\
\end{pmatrix}$, $Q_L=\begin{pmatrix} 
  -1 & 0 & 0 & 0 \\
  0 & 0 & 0 & 0 \\
  0 & 0 & -1 & 0 \\
  0 & 0 & 0 & 0 \\
\end{pmatrix}$. $\textbf{Q}$ has sixty four eigenvalues.\\
The diagonal of $Q_L$ is formed by the charges of four leptons of the same generation, for example $e_L$, $\nu_{eL}$, $e_R$ and $\nu_{eR}$, whereas the diagonal of $Q_Q$ is formed by the charges of a up quark $u$, a down quark $d$, a strange (or a bottom quark $b$) quark $s$ and a charme (or a top quark $t$) quark $c$. For the swap operator $U_{4\otimes 4}$, the four eigenvalues $-1$ are associated to four generations of leptons, the three eigenvalues $+1$ and the one eigenvalue $-1$, are respectively associated to the three colors of left handed quarks and the lepton which form the left handed leptoquark of the Pati-Salam model \cite{ref5}

 \begin{equation*}
\begin{pmatrix}
  u_L^r & u_L^b & u_L^g & \nu_{eL} \\
  d_L^r & d_L^b & d_L^g & e_L \\
  \end{pmatrix}=\begin{pmatrix}
  u_L^r & u_L^b & u_L^g & u_L^w \\
  d_L^r & d_L^b & d_L^g & d_L^w \\
  \end{pmatrix},
  \begin{pmatrix}
  s_L^r & s_L^b & s_L^g & \nu_{\mu L} \\
  c_L^r & c_L^b & c_L^g & \mu_L \\
  \end{pmatrix}=\begin{pmatrix}
  s_L^r & s_L^b & s_L^g & s_L^w \\
  c_L^r & c_L^b & c_L^g & c_L^w \\
  \end{pmatrix} 
\end{equation*}
whereas the other three eigenvalues $+1$ and the one eigenvalue $-1$, are respectively associated to the three colors of right handed quarks and the lepton which form the right handed leptoquark of the Pati-Salam model

\begin{equation*}
\begin{pmatrix}
  u_R^r & u_R^b & u_R^g & \nu_{eR} \\
  d_R^r & d_R^b & d_R^g & e_R \\
  \end{pmatrix}=\begin{pmatrix}
  u_R^r & u_R^b & u_R^g & u_R^w \\
  d_R^r & d_R^b & d_R^g & d_R^w \\
  \end{pmatrix},
  \begin{pmatrix}
  s_R^r & s_R^b & s_R^g & \nu_{\mu R} \\
  c_R^r & c_R^b & c_R^g & \mu_R \\
  \end{pmatrix}=\begin{pmatrix}
  s_R^r & s_R^b & s_R^g & s_R^w \\
  c_R^r & c_R^b & c_R^g & c_R^w \\
  \end{pmatrix} 
\end{equation*}
  
\noindent So, according to \cite{ref5, ref6} we have considered the leptons in the leptoquarks as quarks of color white.

\noindent Finally, the last four eigenvalues $+1$ are associated to four colors of quarks of the third generation, yellow quark added, namely
\begin{equation*}
\begin{pmatrix}
  t_L^r & t_L^b & t_L^g & t_L^y \\
  b_L^r & b_L^b & b_L^g & b_L^y \\
  \end{pmatrix},  \begin{pmatrix}
  t_R^r & t_R^b & t_R^g & t_R^y \\
  b_R^r & b_R^b & b_R^g & b_R^y \\
  \end{pmatrix}, 
\end{equation*}
and that ends the list of sixty four fundamental fermions, right handed neutrinos included.\\
$\Lambda_1=\begin{pmatrix}
  0 & 1 & 0 & 0  \\
  1 & 0 & 0 & 0  \\
  0 & 0 & 0 & 0  \\
  0 & 0 & 0 & 0  \\
\end{pmatrix}$, $\Lambda_2=\begin{pmatrix}
  0 & -i & 0 & 0  \\
  i & 0 & 0 & 0  \\
  0 & 0 & 0 & 0  \\
  0 & 0 & 0 & 0  \\
\end{pmatrix}$, $\Lambda_3=\begin{pmatrix}
  1 & 0 & 0 & 0  \\
  0 & -1 & 0 & 0  \\
  0 & 0 & 0 & 0  \\
  0 & 0 & 0 & 0  \\
\end{pmatrix}$,\\
 $\Lambda_4=\begin{pmatrix}
  0 & 0 & 1 & 0  \\
  0 & 0 & 0 & 0  \\
  1 & 0 & 0 & 0  \\
  0 & 0 & 0 & 0  \\
\end{pmatrix}$, 
$\Lambda_5=\begin{pmatrix}
  0 & 0 & -i & 0  \\
  0 & 0 & 0 & 0  \\
  i & 0 & 0 & 0  \\
  0 & 0 & 0 & 0  \\
\end{pmatrix}$, 
$\Lambda_6=\begin{pmatrix}
  0 & 0 & 0 & 0  \\
  0 & 0 & 1 & 0  \\
  0 & 1 & 0 & 0  \\
  0 & 0 & 0 & 0  \\
\end{pmatrix}$,\\
 $\Lambda_7=\begin{pmatrix}
  0 & 0 & 0 & 0  \\
  0 & 0 & -i & 0  \\
  0 & i & 0 & 0  \\
  0 & 0 & 0 & 0  \\
\end{pmatrix}$, 
$\Lambda_8=\frac{1}{\sqrt{3}}\begin{pmatrix}
  1 & 0 & 0 & 0  \\
  0 & 1 & 0 & 0  \\
  0 & 0 & -2 & 0  \\
  0 & 0 & 0 & 0  \\
\end{pmatrix}$ 
$\Lambda_9=\begin{pmatrix}
  0 & 0 & 0 & 1  \\
  0 & 0 & 0 & 0  \\
  0 & 0 & 0 & 0  \\
  1 & 0 & 0 & 0  \\
\end{pmatrix}$, \\
$\Lambda_{10}=\begin{pmatrix}
  0 & 0 & 0 & -i  \\
  0 & 0 & 0 & 0  \\
  0 & 0 & 0 & 0  \\
  i & 0 & 0 & 0  \\
\end{pmatrix}$, 
$\Lambda_{11}=\begin{pmatrix}
  0 & 0 & 0 & 0  \\
  0 & 0 & 0 & 1  \\
  0 & 0 & 0 & 0  \\
  0 & 1 & 0 & 0  \\
\end{pmatrix}$, 
$\Lambda_{12}=\begin{pmatrix}
  0 & 0 & 0 & 0  \\
  0 & 0 & 0 & -i  \\
  0 & 0 & 0 & 0  \\
  0 & i & 0 & 0  \\
\end{pmatrix}$,\\ 
$\Lambda_{13}=\begin{pmatrix}
  0 & 0 & 0 & 0  \\
  0 & 0 & 0 & 0  \\
  0 & 0 & 0 & 1  \\
  0 & 0 & 1 & 0  \\
\end{pmatrix}$, 
$\Lambda_{14}=\begin{pmatrix}
  0 & 0 & 0 & 0  \\
  0 & 0 & 0 & 0  \\
  0 & 0 & 0 & -i  \\
  0 & 0 & i & 0  \\
\end{pmatrix}$, 
$\Lambda_{15}=\frac{1}{\sqrt{6}}\begin{pmatrix}
  1 & 0 & 0 & 0  \\
  0 & 1 & 0 & 0  \\
  0 & 0 & 1 & 0  \\
  0 & 0 & 0 & -3  \\
\end{pmatrix}$ \\
are the $4\times 4$-Gell-Mann matrices.\\
The formula \cite{ref7} 

\begin{equation*}
U_{4\otimes 4}=\frac{1}{4}\Lambda_0\otimes \Lambda_0+\frac{1}{2}\sum_{i=1}^{15}\Lambda_i\otimes\Lambda_i 
\end{equation*}
yields
\begin{equation*}
\textbf{Q}=\Lambda_0\otimes \Lambda_0\otimes \left(-\frac{1}{12}\Lambda_0-\frac{1}{36}\Lambda_3+\frac{\sqrt{3}}{2}\Lambda_8-\frac{1}{6\sqrt{6}}\Lambda_{15}\right)+\frac{1}{6}\left(\sum_{i=1}^{15}\Lambda_i\otimes\Lambda_i\right)\otimes \Lambda_0
\end{equation*}

\section*{Conclusion}
Thanks to \cite{ref3}, we have an ECO for two leptons and six colored quarks of the SM in one generation. This ECO can be expressed in terms of swap operator $U_{2\otimes 2}$. An ECO for more fermions of the SM in three generations has been obtained in terms of swap operator $U_{3\otimes 3}$. \\
The expression of these ECOs, so the ECO proposed by \cite{ref3}, can be obtained from the relation (\ref{eq13prime}) between the electric charges of leptons and quarks. These expressions allow to say that the eigeinvalues $-1$ of the swap operator are associated to leptons whereas the eigeinvalues $+1$ are asociated to quarks.\\ 
According to the sense taken by an eigenvalue of a swap operator, for obtaining an ECO for all the fermions of the SM, in terms of the swap operator $U_{4\otimes 4}$, we have to introduce fourth generation of leptons, the model of leptoquark of Pati-Salam and quarks of color yellow.

\end{document}